%% file: main.tex
\documentclass{article} 

\usepackage[margin=1in]{geometry}
\PassOptionsToPackage{numbers, compress}{natbib}
\usepackage[]{natbib}
\input{math_commands.tex}

\usepackage[utf8]{inputenc} 
\usepackage[T1]{fontenc}    
\usepackage{amsmath}
\usepackage{amsfonts}

\usepackage{hyperref}
\usepackage{multirow}
\usepackage{url}
\usepackage{subcaption}
\usepackage{graphicx}
\usepackage{colortbl}
\definecolor{green}{rgb}{0.67, 0.88, 0.69}
\definecolor{red}{rgb}{0.97,0.51,0.47}
\definecolor{yellow}{rgb}{1,0.99,0.82}

\title{Loop Unrolled Shallow Equilibrium Regularizer (LUSER) - A Memory-Efficient Inverse Problem Solver}

\author{Peimeng Guan$^1$\thanks{Correspondence: \texttt{guanpeimeng@gatech.edu}} , Jihui Jin$^1$, Justin Romberg$^1$, Mark Davenport$^1$ \\
$^1$Electrical and Computer Engineering\\
Georgia Institute of Technology \\
Atlanta, GA 30332, United States\\
}

\begin{document}
\graphicspath{{figs/}}

\maketitle

\begin{abstract}
In inverse problems we aim to reconstruct some underlying signal of interest from potentially corrupted and often ill-posed measurements. Classical optimization-based techniques proceed by optimizing a data consistency metric together with a regularizer. Current state-of-the-art machine learning approaches draw inspiration from such techniques by unrolling the iterative updates for an optimization-based solver and then learning a regularizer from data. This \emph{loop unrolling} (LU) method has shown tremendous success, but often requires a deep model for the best performance leading to high memory costs during training. Thus, to address the balance between computation cost and network expressiveness, we propose an LU algorithm with shallow equilibrium regularizers (LUSER). These implicit models are as expressive as deeper convolutional networks, but far more memory efficient during training. The proposed method is evaluated on image deblurring, computed tomography (CT), as well as single-coil Magnetic Resonance Imaging (MRI) tasks and shows similar, or even better, performance while requiring up to $8 \times$ less computational resources during training when compared against a more typical LU architecture with feedforward convolutional regularizers.
\end{abstract}

\section{Introduction}
In an inverse problems we face the task of reconstructing some data or parameters of an unknown signal from indirect observations. The forward process, or the mapping from the data to observations, is typically well known, but ill-posed or non-invertible. More formally, we consider the task of recovering some underlying signal $\vx$ from measurements $\vy$ taken via some forward operator $\mA$ according to 
\begin{equation}\label{eq:1}
    \vy = \mA\vx + \boldsymbol{\eta},
\end{equation}
where $\boldsymbol{\eta}$ represents noise. The forward operator can be nonlinear, but to simplify the notation, we illustrate the idea in linear form throughout this paper.
A common approach to recover the signal is via an iterative method based on the least squares loss:
\begin{equation}\label{eq:lsq}
    \widehat{\vx} = \argmin_{\vx}~\norm{\vy-\mA\vx}^2.
\end{equation}
For many problems of interest, $\mA$ is ill-posed and does not have full column rank. Thus, attempting to solve \eqref{eq:lsq} does not yield a unique solution. To address this, we can extend \eqref{eq:lsq} by including a regularizing term to bias the inversion towards solutions with favorable properties. Common examples of regularization include $\ell_2$, $\ell_1$, and Total Variation (TV). Each regularizer encourages certain properties on the estimated signal $\widehat\vx$ (e.g., smoothness, sparsity, piece-wise constant, etc.) and is often chosen based on task-specific prior knowledge.

Recent works \cite{ongie2020deep} attempt to tackle inverse problems using more data-driven methods. Unlike typical supervised learning tasks that attempt to learn a mapping purely from examples, deep learning for inverse problems have access to the forward operator and thus should be able to guide the learning process for more accurate reconstructions. One popular approach to incorporating knowledge of the forward operator is termed \emph{loop unrolling} (LU). These methods are heavily inspired by standard iterative inverse problem solvers, but rather than use a hand tuned regularizer, they instead learn the update with some parameterized model. They tend to have a fixed number of iterations (typically around 5-10) due to computational constraints. \cite{gilton2021deep} proposes an interesting alternative that takes advantage of \emph{deep equilibrium} (DEQ) models \cite{bai2019deep, bai2020multiscale, fung2021fixed, el2021implicit} that we refer to as DEQ4IP. Equilibrium models are designed to recursively iterate on their input until a ``fixed point" is found (i.e., the input no longer changes after passing through the model). They extend this principle to the LU method, choosing to iterate until convergence rather than for a ``fixed budget". 

\paragraph{Our Contributions.} 
We propose an alternative novel architecture for solving inverse problems called \emph{Loop Unrolled Shallow Equilibrium Regularizer} (LUSER). It incorporates knowledge of the forward model by adopting the principles of LU architectures while reducing its memory consumption by using a shallow (relative to existing feed-forward models) DEQ as the learned regularizer update. Unlike DEQ4IP that converts the entire LU architecture into a DEQ model, we only convert the learned regularizer at each stage. This has the advantage of simplifying the learning task for DEQ models, which can be unstable to train in practice. To our knowledge, this is the first use of multiple sequential DEQ models within a single architecture for solving inverse problems.   Our proposed architecture (\emph{i}) reduces the number of forward/adjoint operations compared to the work proposed by \cite{gilton2021deep}, (\emph{ii}) reduces the memory footprint during training without loss of expressiveness as demonstrated by our experiments, and (\emph{iii}) is more stable to train in practice than DEQ4IP. We empirically demonstrate better reconstruction across multiple tasks than state-of-the-art LU alternatives with a similar number of parameters, with the ability to reduce computational memory costs during training by a factor of up to $8\times$.

The remainder of the paper is organized as follows. Section \ref{sec:related works} reviews related works in solving inverse problems. Section \ref{sec: methodology} introduces the proposed LUSER, which we compare with other baseline methods in image deblurring, CT, and MRI tasks in Section \ref{sec: experiment}. We conclude in Section \ref{sec: conclusion} with a brief discussion.

\section{Related Work} \label{sec:related works}

\subsection{Loop Unrolling}
As noted above, a common approach to tackling an inverse problem is to cast it as an optimization problem consisting of the sum of a data consistency term and a regularization term
\begin{equation}\label{eq:2}
    \min_{x}~\|\vy-\mA\vx\|^2_2 + \gamma \, r(\vx),
\end{equation}
where $r$ is a regularization function mapping from the domain of the parameters of interest to a real number and $\gamma \geq 0$ is a well-tuned coefficient. The regularization function is chosen for specific classes of signals to exploit any potential structure, e.g., $\| \vx\|_2$ for smooth signals and $\|\vx\|_0$ or $\| \vx \|_1$ for sparse signals. The total-variation (TV) norm is another popular example of a regularizer that promotes smoothness while preserving edges and is often used for 

When $r$ is differentiable, the solution of \eqref{eq:2} can be obtained in an iterative fashion via gradient descent. For some step size $\lambda$ at iteration $k = 1,2,\ldots, K$, we apply the update:
\begin{equation}\label{eq: GD}
    \vx_{k+1} = \vx_{k} + \lambda \mA^\top (\vy-\mA\vx_{k}) - \lambda \gamma \nabla r(\vx_{k}).
\end{equation}
For non-differentiable $r$, the more generalized \emph{proximal gradient} algorithm can be applied with the following update, where $\tau$ is a well-tuned hyperparameter related to the proximal operator: 
\begin{equation}
    \label{eq:update}
    \vx_{k+1} = \mbox{prox}_{\tau, r}(\vx_{k} + \lambda \mA^\top (\vy-\mA\vx_{k})).    
\end{equation}
The \emph{loop unrolling} (LU) method considers running the update in \eqref{eq: GD} or \eqref{eq:update}, but replaces $\lambda \gamma \nabla r$ or the proximal operator with a learned neural network instead. The overall architecture repeats the neural network based update for a pre-determined number of iterations, fixing the overall computational budget. Note that the network is only implicitly learning the regularizer. In practice, it is actually learning an update step, which can be thought of as de-noising or a projection onto the manifold of the data. LU is typically trained end-to-end, i.e., when fed some initialization $\vx_0$, the network will output the final estimate $\vx_K$, compute a loss with respect to the ground truth, and then back-propagate across the entire computational graph to update network parameters. While end-to-end training is easier to perform and encourages faster convergence, it requires all intermediate activations to be stored in memory. Thus, the maximum number of iterations is always kept small compared to classical iterative inverse problem solvers. 

Due to the limitation in memory, there is a trade-off between the depth of a LU and the richness of each regularization network. Intuitively, one can raise the network performance by increasing the number of loop unrolled iterations. For example, \cite{gilton2021deep} extends the LU model to potentially infinite number of iterations using an implicit network, and \cite{https://doi.org/10.48550/arxiv.1911.10914} allows deeper unrolling iterations using invertible networks, while requiring recalculation of the intermediate results from the output in training phase. This approach can be computationally intensive for large-scale inverse problems or when the forward operator is nonlinear and computationally expensive to apply. For example, the forward operator may involve solving differential equations such as the wave equation for seismic wave propagation \cite{chapman2004fundamentals} and the Lorenz equations for atmospheric modeling\cite{oestreicher2022history}.

Alternatively, one can design a richer regularization network. For example \cite{fabian2022humus} uses a transformer as the regularization network and achieves extremely competitive results in the fastMRI challenge \cite{https://doi.org/10.48550/arxiv.1811.08839}, but requires multiple 24GB GPU for training with batch size of 1, which is often impractical, especially for large systems. Our design strikes a balance in the expressiveness in regularization networks and memory efficiency during training. Our proposed work is an alternative method to achieve a rich regularization network without the additional computational memory costs during training.

\subsection{Deep Equilibrium Models for Inverse Problems (DEQ4IP)}
\emph{Deep equilibrium} (DEQ) models introduce an alternative to traditional feed-forward networks \cite{bai2019deep, bai2020multiscale, fung2021fixed, el2021implicit}. Rather than feed an input through a fixed (relatively small) number of layers, DEQ models solve for the ``fixed-point" given some input. More formally, given a network $f_\theta$ and some input $\vx^{(0)}$ and $ \vy$, we recursively apply the network via
\begin{equation}
    \vx^{(k+1)} = f_\theta(\vx^{(k)}, \vy),
\end{equation}
until convergence.\footnote{Note that, since our approach will ultimately use both methods, to aid in a clearer presentation we use subscript, i.e., $\vx_k$, to denote the LU iterations, and superscript with parenthesis, i.e., $\vr^{(i)}$, to denote the iterations in the deep equilibrium model.} In this instance, $\vy$ acts as an input injection that determines the final output. This is termed the forward process. The weights $\theta$ of the model can be trained via implicit differentiation, removing the need to store all intermediate activations from recursively applying the network. This allows for deeper, more expressive models without the associated memory footprint to train such models. 

\cite{gilton2021deep} demonstrates an application of one such model, applying similar principles to a single iteration of an LU architecture. Such an idea is a natural extension as it allows the model to ``iterate until convergence" rather than rely on a ``fixed budget". In another word, looping over \eqref{eq: GD} and \eqref{eq:update} many times (in practice, usually around 50 iterations) until $\vx_k$ converges.  However, such a model can be unstable to train and often performs best with pre-training of the learned portion of the model (typically acting as a learned regularizer/de-noiser). It is also important to note is that such a model would have to apply the forward operator (and potentially the adjoint) many times during the forward process. Although this can be accelerated to reduce the number of applications, it is still often more than the number of applications for an LU equivalent. This can be an issue if the forward operator is computationally expensive to apply, an issue LU methods somewhat mitigate by fixing the total number of iterations.

\subsection{Alternative Approaches to Tackle Memory Issues}
I-RIM \cite{https://doi.org/10.48550/arxiv.1911.10914} is a deep invertible network that address the memory issue by recalculating the intermediate results from the output. However it is not ideal when forward model is computationally expensive. Gradient checkpointing \cite{sohoni2019low} is another practical technique to reduce memory costs for deep neural networks. It saves intermediate activations of some checkpoint nodes, and recomputes the forward pass between two checkpoints for backpropagation. However, it is not an easy and efficient technique to implement for a weight-tied neural network.

\section{Methodology} \label{sec: methodology}
LU methods have been the state-of-the-art approach for solving inverse problems due to their stability when training, inclusion of the forward model, and their near instantaneous inference times. However, there is a noticeable trade-off in terms of the memory requirements when training these models and the accuracy for even medium scale problems, making them nearly impossible to use for much larger scale inverse problems. DEQ4IP offers an interesting alternative, drastically reducing the memory required during training and allowing the flexibility to adjust accuracy during reconstruction when performing inference. However, these models can potentially suffer when the forward/adjoint operators are computationally expensive, particularly when it takes more iterations to converge than a standard fixed LU method. 

To address these concerns, we propose a novel architecture for solving inverse problems called \emph{Loop Unrolled Shallow Equilibrium Regularizer} (LUSER) as an approach that limits the number of forward/adjoint calls for particularly complex inverse problems while drastically reducing the memory requirements for training, allowing us to scale up to much larger inverse problems (or require less GPU memory for existing problems). LUSER achieves this by adopting DEQ models as the learned regularizer update in a standard LU architecture. The implicit DEQ models are smaller in size but just as expressive as typical convolutional models used as the learned regularizer update allowing for an accurate reconstruction with less computational memory costs. Furthermore, learning a proximal update is a far simpler task compared to solving the inverse problem as a whole, reducing the instability during training commonly faced by DEQ models. 

We adopt a ``proximal gradient descent" styled LU architecture. The network takes in measurements $\vy$ and some initial estimate $\vx_0$. The architecture consists of $K$ stages, alternating between a gradient descent step $\vd_k = \vx_{k} + \lambda \mA^\top (\vy-\mA\vx_{k})$ for $k = 1,2, \ldots, K$, followed by a feed-forward pass of the shallow equilibrium model acting as a proximal update block. 

\begin{figure}[htp]
    \centering
    \includegraphics[scale=0.45]{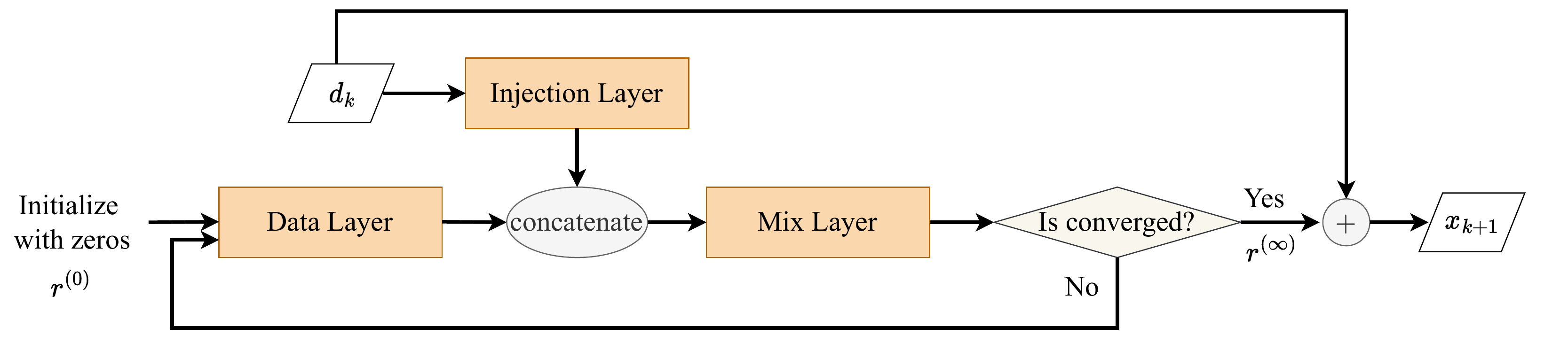}
    \caption{A proximal block in LUSER}
    \label{fig:luser}
\end{figure}

Figure \ref{fig:luser} illustrates a single block in LUSER as a learned proximal operator. The input $\vd_k$ from the previous stage is processed only once by a set of input injection layers to avoid redundant computation. The input injection will determine the final fixed point output \cite{fung2021fixed}. The recursive portion of the proximal block consists of two sets of layers: the data layer and the mixing layer. The current estimate of the fixed point is first passed through the data layer before being concatenated with the input injection and processed by the mixing layer. This process repeats until a fixed point is found. 
Although Figure \ref{fig:luser} shows the simplest way of finding the fixed point, in practice acceleration or other fixed point solvers are applied to solve for the fixed point. We apply Anderson acceleration \cite{walker2011anderson} for all of our models. 

At each stage, the shallow equilibrium regularizer attempts to output its best estimate of the ground truth $\vx^*$. We introduce a skip connection between the input $\vd_k$ and the final output so that the regularizer is learning the residual $\vr$ between the input $\vd_k$ and the ground truth $\vx^*$ instead. When $\vx_k$ converges to $\vx^*$, $\vd_k$ will also converge to $\vx^*$ and we expect the residual to be closer and closer to the zero vector. Therefore, we initialize the input of data layer with a zero vector with the same dimensions as the input $\vd_k$ in the hopes that fewer iterations will be needed at later stages as the current estimate $\vx_k$ converges towards the ground truth. Let $\vr^{(i)}$ denote the $i^{th}$ update of the residual, where $\vr^{(0)} = \boldsymbol{0}$ and $\vr^{(\infty)}$ be the fixed point solution of the residual. Let $\oplus$ denote the concatenation operator. The process in one loop unrolled block can be formulated as the following:
\begin{equation}
    \vx_{k+1} = \vd_k+ \text{MixLayer}\left(\text{DataLayer}(\vr^{(\infty)}) \oplus \text{InjectLayer}(\vd_k)\right).
\end{equation}
Because of the fixed point solver within each regularization block, LUSER uses fewer layers to achieve a similar level of performance, thus we call LUSER a shallow equilibrium regularizer.

We also explore two variants of LUSER, dubbed LUSER-SW and LUSER-DW. LUSER-SW refers to a shared-weight version of the proposed algorithm, where the proximal operator in all loop unrolled iterations are weight-tied and thus identical. Theoretically, the regularizer should be able to handle any input regardless of the iteration step. However, in practice, the distribution of intermediate reconstructions can be vastly different. Thus, training a single model to handle all these instances can be a daunting task, leading to poor generalization across the different stages. Since the total computational budget is fixed, one approach can be to use different weights (DW) for the learned proximal operator at each stage to handle the potentially different distributions. This will increase the number of parameters that need to be trained and stored, but since LUSER already has so few parameters to begin with, expanding to the different weight variant is still feasible. This paper will compare two variants of LUSER with other architectures in different tasks.

DEQ4IP relies on the learned proximal operator to have the same weights, thus we cannot include a different weight variant for comparison. On the other hand, the same principles can be applied to the LU variant of DEQ4IP. However, we aim to only compare models of similar number of parameters (or less in the case of LUSER-SW), and thus restrict our attentions to the shared weight variant of LU only.

\section{Experiments and Results}\label{sec: experiment}
In this section, we compare our proposed networks to LU with DnCNN as proximal operator (LU-DnCNN) and DEQ4IP on three different tasks: image deblurring, computed tomography (CT), and single-coil accelerated Magnetic Resonance Imaging (MRI). The datasets we use are RGB CelebA  \cite{liu2015faceattributes}, LoDoPaB-CT \cite{leuschner2021lodopab}, and single-coil knee data from fastMRI \cite{https://doi.org/10.48550/arxiv.1811.08839} respectively. We also experiment with incorporating an auxiliary loss of including an MSE loss on intermediate reconstructions with the ground-truth instead of just the final output. This is done for the LU models (LU-DnCNN, LUSER).

\subsection{Experimental Setup}
In the image deblurring task, the blurry image is obtained by applying a $(9\times 9)$ Gaussian kernel with variance of $5$ to an image with additive white Gaussian noise with a standard deviation of 0.01. If the image is RGB, the same kernel is applied to all channels. For accelerated MRI tasks, measurements in k-space (or frequency domain) are often subsampled due to the cost in measurement. The goal of MRI reconstruction is to recover the underlying physical structure from subsampled noisy measurements. We simulate the forward operator $\mA$ with a 2-dimensional Fourier transform with randomly selected rows. We consider two common subsampling scenarios: 4x and 8x acceleration, or subsampling the columns in full measurement by a factor of 4 and 8 respectively. For the CT task, the forward operator is a Radon transform, and we uniformly select 200 out of 1000 angles in measurements. The adjoint of measurement $\mA^\top \vy$ is used as initialization for MRI and CT tasks, which brings the measurement back to the signal domain. However, in the deblurring task, since the measurement lies in a same domain as the underlying clean image, $\vy$ is used as the initial guess. Notice that some works use the filtered backprojection as the initialization for CT, such as \cite{khorashadizadeh2022conditional}, but we use the adjoint for the purpose of consistency.

\begin{table}[htp]
    \centering
    \caption{\label{tab:LUSER} Architecture of Proximal Network in LUSER}
    \begin{tabular}{ c|c } 
          & Layer Details \\ \hline
         \multirow{2}{*}{Injection Layer} &  SN(conv($C_{\text{in}}$:1,  $C_{\text{out}}$:32, $\text{ks}$:3, $\text{pad}$:1)) + LeakyReLU\\
          & SN(conv($C_{\text{in}}$:32,  $C_{\text{out}}$:32, $\text{ks}$:3, $\text{pad}$:1)) + LeakyReLU\\ \hline
         \multirow{2}{*}{Data Layer} &  SN(conv($C_{\text{in}}$:1,  $C_{\text{out}}$:32, $\text{ks}$:3, $\text{pad}$:1)) GN + LeakyReLU\\
          & SN(conv($C_{\text{in}}$:1,  $C_{\text{out}}$:32, $\text{ks}$:3, $\text{pad}$:1)) GN + LeakyReLU\\ \hline
         \multirow{3}{*}{Mix Layer} &  SN(conv($C_{\text{in}}$:64,  $C_{\text{out}}$:64, $\text{ks}$:3, $\text{pad}$:1)) GN + LeakyReLU\\
          & SN(conv($C_{\text{in}}$:64,  $C_{\text{out}}$:64, $\text{ks}$:3, $\text{pad}$:1)) GN + LeakyReLU\\
          & SN(conv($C_{\text{in}}$:64,  $C_{\text{out}}$:1, $\text{ks}$:3, $\text{pad}$:1))
    \end{tabular}
\end{table}
We fix the budget of LU-DnCNN and LUSER to be a total of 8 iterations, while as we allow DEQ4IP to iterate until it reaches a fixed point. We use a DnCNN adopted from \cite{ryu2019plug} with 17 convolutional layers with 64 channels, followed by BatchNorm and ReLU activations for the regularizer for LU-DnCNN and DEQ4IP. For the learned regularizer update in LUSER, we use 2 convolutional layers each for the input injection layer and data layer. The mixing layer contains 3 convolutional layers. 

In order to stablize training for DEQ inspired models, we wrap all convolutional layers in DEQ4IP and LUSER with Spectral Norm (SN) \cite{miyato2018spectral}. We list more details for the learned proximal network for LUSER in Table \ref{tab:LUSER} for the case when the input has a single channel. $C_{\text{in}}$ and $C_{\text{out}}$ denote the input and output channels, $\text{ks}$ refers to the kernel size of a convolutional layer, $\text{pad}$ denotes the padding in 2-dimension, and GN stands for GroupNorm.

We use two metrics to evaluate the quality of reconstruction: Peak Signal-to-Noise Ratio (PSNR) in dB and the Structural Similarity Index (SSIM): 
\begin{equation}
    \text{PSNR}(\vx, \widehat{\vx}) = 20 \log_{10} \left( \frac{\max(\vx)}{\text{MSE}(\vx, \widehat{\vx})}\right)
\end{equation}
\begin{equation}
    \text{SSIM}(\vx, \widehat{\vx}) = \frac{(2\mu_{\vx}\mu_{\widehat{\vx}} + C_1)(2 \sigma_{\vx \widehat{\vx}} + C_2)}{(\mu_{\vx}^2 + \mu_{\widehat{\vx}}^2 + C_1)(\sigma_{\vx}^2 + \sigma_{\widehat{\vx}}^2 + C_2)}
\end{equation}
where $\text{MSE}(\vx, \widehat{\vx}) = \mathbb{E}[ \|\vx-\widehat{\vx} \|^2]$ is the mean squared error. In SSIM, $\mu_{\vx}$ and $\mu_{\widehat{\vx}}$ are pixel means of $\vx$ and $\widehat{\vx}$, $\sigma_{\vx}^2$ and $\sigma_{\widehat{\vx}}^2$ are variances of $\vx$ and $\widehat{\vx}$, and $\sigma_{\vx \widehat{\vx}}$ is their covariance. $C_1$ and $C_2$ are small positive numbers to stabilize SSIM. Note that although we use the same models as \cite{gilton2021deep}, we train our models from scratch and report lower values on the MRI task. We suspect that this is due to evaluating with a single channel only. When we include the imaginary channel (for a total of 2 channels), the metrics we recorded are more aligned with those reported in \cite{gilton2021deep}. 

All models are trained with a single RTX6000 24GB GPU.

\subsection{Main Results}
Table \ref{tab:full_result} compares the average PSNR and SSIM. The different weight version of LUSER outperforms LU-DnCNN with a similar number of network parameters. LUSER-SW achieves similar level of performance in most tasks with only 5 layers, versus 17 layers in LU-DnCNN. DEQ4IP attains the best performance in image delurring task, but it requires repetitive computation of forward/adjoint operators. Training shared-weight architectures with auxillary losses improves the reconstruction quality in most tasks.

\begin{table}[h]
\caption{Average PSNR and SSIM for test set, the best two performances are in bold.}
\label{tab:full_result}
\begin{center}
\begin{tabular}{c|ccccccc}
\multicolumn{1}{c|}{PSNR} &\multicolumn{2}{c}{\bf LU-DnCNN} &\multicolumn{1}{c}{\bf DEQ4IP } &\multicolumn{2}{c}{\bf LUSER-SW} &\multicolumn{2}{c}{\bf LUSER-DW}  \\
 SSIM & Final loss & Aux loss & Final loss & Final loss & Aux loss  & Final loss & Aux loss \\ 
\hline
\multirow{2}{*}{CelebA}   & 29.93 & 30.39 & \textbf{31.57} & 30.30 & 30.65 & \textbf{31.40} & 31.15  \\
         & 0.862 & 0.871 & \textbf{0.895} & 0.869 & 0.878 & \textbf{0.891} & 0.888  \\ \hline
                         
\multirow{2}{*}{CT}   & 30.59 & 31.59 & \textbf{31.79} & 28.82 & 28.04 & \textbf{31.83} & 31.66  \\
     & 0.844 & 0.859 & \textbf{0.868} & 0.801 & 0.797 & \textbf{0.860} & 0.859  \\ \hline
                         
\multirow{2}{*}{4x MRI}   & 29.01 & 29.02 & 29.01 & 28.82 & 29.18 & \textbf{29.86} & \textbf{29.37}  \\
         & 0.668 & 0.671 & 0.678 & 0.662 & 0.685 & \textbf{0.740} &\textbf{0.713}  \\ \hline
                         
\multirow{2}{*}{8x MRI}     & 27.50 & \textbf{27.65} & 27.51 & 27.42 & 27.42 & \textbf{28.06} & 27.55  \\
           & 0.576 & 0.572 & 0.570 & 0.562 & 0.560 & \textbf{0.630} & \textbf{0.596}  
\end{tabular}
\end{center}
\end{table}
\begin{figure}
    \centering
    \includegraphics[width=\textwidth]{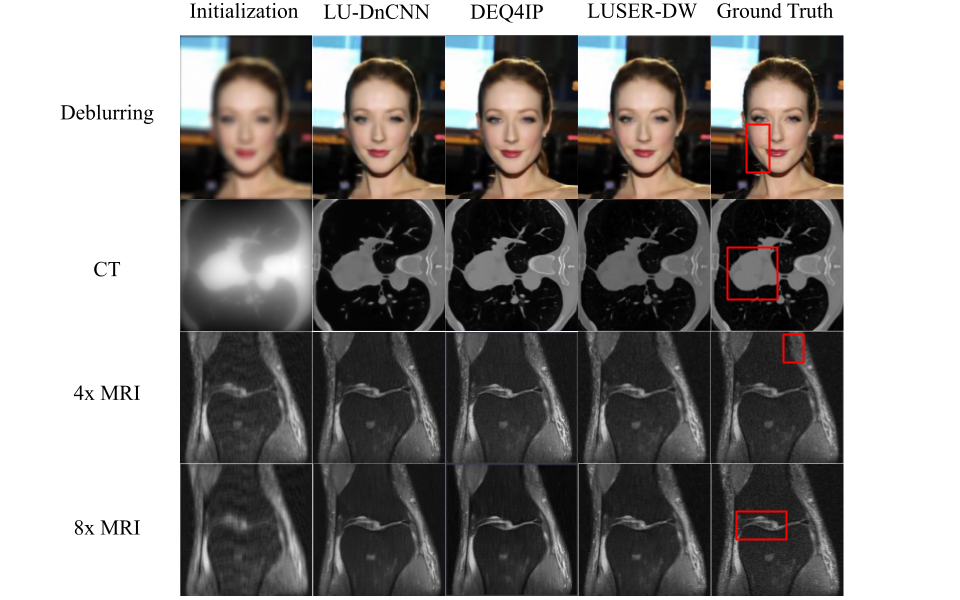}
    \caption{Representative reconstructions, where each row represents one task. The left-most column shows the initialization or input to the networks; the middle three columns show the reconstructions for LU-DnCNN, DEQ4IP and LUSER-DW; and the right-most column shows the underlying true images. Regions corresponding to qualitative improvements are emphasized in red boxes in the last column.}
    \label{fig:res_imgs}
\end{figure}
Figure \ref{fig:res_imgs} shows representative reconstruction results. The overall performances are similar, but LUSER-DW attains better qualities in detailed structures, especially compared to LU-DnCNN. The areas with improvements are emphasized with red boxes in the ground truth images.

\begin{table}[h]
\caption{Comparison of network sizes and maximum possible batch sizes during training. Entries with `-' denote that the architecture with a batch size of 1 cannot fit into a particular GPU RAM capacity.}
\label{tab:memory}
\begin{center}
\begin{tabular}{c|ccccc}
\multicolumn{1}{c|}{} & \multicolumn{1}{c}{GPU RAM} &\multicolumn{1}{c}{\bf LU-DnCNN} &\multicolumn{1}{c}{\bf DEQ4IP} &\multicolumn{1}{c}{\bf LUSER-SW} &\multicolumn{1}{c}{\bf LUSER-DW}  \\
 Max Batch Size & Capacity & (17 layers) & (17 layers) & (5 layers)  & (5 layers) \\ 
\hline
\multirow{4}{*}{deblurring}  &\#Params & 558,580 & 558,580   & 96,503    & 770,073 \\
                                     & 8 GB  & 1 & 16    & 10  & 10 \\
                                     & 10 GB & 1 & 16    & 14  &  14 \\
                                     & 24 GB & 4 & 68   & 34  & 34 \\ \hline
\multirow{4}{*}{CT}  & \#Params & 556,033 & 556,033   & 93,954    & 751,625 \\
                             & 8 GB  & - & 4    & 2     & 2 \\
                             & 10 GB & - & 6     & 4     & 4 \\
                             & 24 GB & 2 & 20    & 10    & 10 \\ \hline
\multirow{4}{*}{MRI}  & \#Params & 557,185 & 557,185   & 95,107    & 760,849 \\
                             & 8 GB  & - & 4     & 4     & 4 \\
                             & 10 GB & 2  & 6    & 4     & 4  \\
                             & 24 GB & 4 & 16    & 12    & 12
\end{tabular}
\end{center}
\end{table}

Table \ref{tab:memory} compares the network sizes (number of parameters) as well as the maximum training batch sizes for three commonly seen GPU RAM capacities: 8 GB, 10 GB and 24 GB. Batch sizes are recorded with maximum even numbers, except when it is 1 for stochastic gradient descent. Notice that in MRI, because the Fourier transformation is implemented in tensor form, the minimum batch size it can take is 2. We use the batch size as a proxy for the memory requirements during training. Since DEQ4IP is an extension of LU-DnCNN, their networks are of the same size, but implicit DEQ models support larger batch sizes making DEQ4IP far more memory-efficient during training. The advantages of using DEQ models for LUSER and DEQ4IP are particularly highlighted in the case of limited memory (smaller GPUs). LU-DnCNN is unable to even train for the CT and MRI task with limited memory constraints, while LUSER and DEQ4IP can. This pattern is expected to repeat for more large scale tasks where standard LU architectures will be unable to train at all due to memory requirements. It is important to note that memory requirements depend more on the depth of the network than the number of parameters. For example, even though LUSER-DW and LUSER-SW have different numbers of parameters, they share the same architecture/depth and thus use roughly the same amount of memory during training.

We observe that LUSER takes longer to train in each epoch than LU-DnCNN, but the training converges in fewer iterations than LU-DnCNN. The total training time for all methods are roughly the same for the image deblurring and MRI tasks, but DEQ4IP takes longer to train on the CT problem as it is finding a fixed-point with a more complicated forward operator. During evaluation, LU-DnCNN is the fastest, and the DEQ4IP and LUSER are in similar speed. When the forward operator is computationally intensive, LUSER evaluates faster than DEQ4IP. Table \ref{tab:comp} compares the various properties among the three architectures.

\begin{table}[htp]
\caption{Method Comparisons}
\label{tab:comp}
\centering
\renewcommand{\arraystretch}{1.3}
\begin{tabular}{|c|c|c|c|}
    \hline
     & LU & DEQ4IP & LUSER \\ \hline
    Training Time & \cellcolor{green} Fast & \cellcolor{red} Slow & \cellcolor{yellow} Moderate  \\ \hline
    Evaluation Time & \cellcolor{green} Fast & \cellcolor{red} Slow & \cellcolor{yellow} Moderate \\ \hline
    Training Stability & \cellcolor{green} Stable & \cellcolor{red} Unstable & \cellcolor{green} Stable\\ \hline
    Network Size & \cellcolor{red} Large & \cellcolor{red} Large & \cellcolor{green} Small \\ \hline
    Expressiveness & \cellcolor{red} Low & \cellcolor{yellow} Moderate & \cellcolor{green} High \\ \hline
    Training Memory Usage & \cellcolor{red} Large & \cellcolor{green} Small & \cellcolor{yellow} Moderate \\
    \hline
\end{tabular}

\end{table}

\subsection{Effect of number of Layers in LUSER-SW}
The experiments above show the results of LUSER-SW with only 5 layers as extreme case to highlight the effectiveness of even a vastly smaller network. The number of layers typically correlate with the expressiveness of a deep feedforward neural network. We also explore the effect of number of layers in the reconstruction quality for the learned proximal operators of LUSER. Table \ref{tab:layer_result} shows the average PSNR and SSIM with one extra layer in the Mix Layer. It raises the PSNR by 1 dB and SSIM by 0.2, but sacrifices some memory in training. This suggests that one can design the network in LUSER according to their need in reconstruction quality and memory constraints.

\begin{table}[h]
\caption{Average PSNR and SSIM for LUSER-SW with more layers on CT task.}
\label{tab:layer_result}
\begin{center}
\begin{tabular}{c|cccc}
\multicolumn{1}{c|}{PSNR} &\multicolumn{2}{c}{\bf 5 layers} &\multicolumn{2}{c}{\bf 6 layer}  \\
 SSIM & Final loss & Aux loss & Final loss & Aux loss  \\ 
\hline
\multirow{2}{*}{CT}   & 28.82 & 28.04 & 29.86 & 29.04  \\
                      & 0.801 & 0.797 & 0.823 & 0.808  \\
\end{tabular}
\end{center}
\end{table}

\subsection{Effect of weight sharing}
As shown in Table \ref{tab:full_result}, LUSER-DW has better performance than LUSER-SW with equivalent block structure, but has a larger number of parameters. We also explore the possibility of reusing multiple LUSER proximal blocks across loop unrolled iterations. For example, in MRI tasks, we repeat 4 different proximal blocks over 8 iterations. Let $g_i$ denote the $i^{th}$ proximal block where $i= \{1,2,3,4\}$, and we form the LUSER network with the sequence of proximal blocks in order of $g_1, g_1, g_2, g_2, g_3, g_3, g_4, g_4$. Now, the number of parameters is only half of that in LUSER-DW, but maintains a similar level of performance. We denote this variant LUSER-PSW, which stands for partially shared-weight. Table \ref{tab:PSW} compares the average PSNR and SSIM of LUSER-SW, LUSER-PSW and LUSER-DW with the same block structure, which are trained using final loss only.

\begin{table}[h]
    \centering
    \caption{Average PSNR and SSIM for LUSER with different weight-sharing schematics}
    \label{tab:PSW}    
    \begin{tabular}{c|ccc}
        PSNR & \multirow{2}{*}{LUSER-SW} & \multirow{2}{*}{LUSER-PSW} & \multirow{2}{*}{LUSER-DW}  \\
        SSIM & & & \\ \hline
        \multirow{2}{*}{4x MRI} & 28.82 & 28.85 & 29.86 \\
            & 0.66 & 0.74 & 0.74\\ \hline
        \multirow{2}{*}{8x MRI} & 27.42 & 28.04 & 28.06 \\
            & 0.56 & 0.63 & 0.63\\ 
    \end{tabular}
\end{table}
\section{Conclusion}\label{sec: conclusion}
Loop unrolling architectures with deep convolutional layers as the learned regularizer update are a popular approach for solving inverse problems. Although its variants achieve state-of-the-art
results across a variety of tasks, LU algorithms incur a huge memory cost when training due to the requirement of saving all intermediate activations, sometimes even requiring multiple GPUs to train on complex tasks.  DEQ4IP offers an interesting alternative via extending LU to infinitely many layers by finding a fixed point solution, but can be impractical when the forward/adjoint operators are nonlinear or larger in scale. 
To address the memory issue, we proposed two variants of loop unrolling architectures with deep equilibrium models as the learned regularizer updates, LUSER-SW and LUSER-DW. We verify the memory savings (by comparing batch sizes) relative to loop unrolling algorithms with a DnCNN model. Across all tasks, LUSER-DW outperforms LU-DnCNN with a similar number of network parameters, while reducing the memory requirements by a factor of 5 or more. LUSER offers a path forward for large-scale, complex, non-linear inverse problems that are currently infeasible to train with existing methods.

\section*{Acknowledgment}
Guan and Davenport are supported by the Center for Energy and Geo Processing (CeGP) at Georgia Tech and the King Fahd University of Petroleum and Minerals. Jin and Romberg are supported in part by the Office of Naval Research Task Force Ocean under Grant No. N00014-19-1-2639.

\bibliographystyle{unsrt}
\bibliography{reference.bib}


\end{document}

%% file: math_commands.tex
\usepackage{amsmath,amsfonts,bm}

\newcommand\norm[1]{\lVert#1\rVert}

\def\1{\bm{1}}

\def\vd{{\bm{d}}}

\def\vr{{\bm{r}}}

\def\vx{{\bm{x}}}
\def\vy{{\bm{y}}}

\def\mA{{\bm{A}}}

\DeclareMathAlphabet{\mathsfit}{\encodingdefault}{\sfdefault}{m}{sl}
\SetMathAlphabet{\mathsfit}{bold}{\encodingdefault}{\sfdefault}{bx}{n}

\DeclareMathOperator*{\argmin}{arg\,min}